\documentclass[reprint,nofootinbib,amsmath,amssymb,fontenc aps]{revtex4-1}
\usepackage{graphicx,dcolumn,bm,hyperref,float}
\usepackage{anyfontsize}
\usepackage{color}
\usepackage{xcolor}

\newcommand{\beq}{\begin{equation}}
\newcommand{\eeq}{\end{equation}}
\newcommand{\bea}{\begin{eqnarray}}
\newcommand{\eea}{\end{eqnarray}}
\newcommand{\bef}{\begin{figure}}
\newcommand{\eef}{\end{figure}}

\newcommand{\mpl}{M_{\mbox{\tiny{Pl}}}}
\newcommand{\tpl}{t_{\mbox{\tiny{Pl}}}}
\newcommand{\pr}{\mathcal{P}}
\newcommand{\CMB}{\mbox{CMB}}
\newcommand{\tCMB}{\mbox{\tiny{CMB}}}
\newcommand{\eq}{\mbox{eq}}
\newcommand{\teq}{\mbox{\tiny{eq}}}

\begin{document}

\title{Constraints on an Anisotropic Universe }

\author{Mark P.~Hertzberg$^{1,2}$}
\email{mark.hertzberg@tufts.edu}
\author{Abraham Loeb$^2$}
\email{aloeb@cfa.harvard.edu}
\affiliation{$^1$Institute of Cosmology, Department of Physics and Astronomy, Tufts University, Medford, MA 02155, USA
\looseness=-1}
\affiliation{$^2$Department of Astronomy, Harvard University, 60 Garden Street, Cambridge, MA 02138, USA
\looseness=-1}

\begin{abstract}
We analyze the possibility of global anisotropy of the universe. We consider an altered Friedmann Lemaitre Robertson Walker metric in which there are different scale factors along the three different axes of space. We construct the corresponding altered Friedmann equations. We show that any initial anisotropies decrease into the future. At late times, the difference in Hubble parameters changes as $1/\sqrt{t}$ in a radiation dominated era and as $1/t$ in a matter dominated era. We use constraints from Big Bang Nucleosynthesis and the Cosmic Microwave Background  to constrain the level of anisotropies at early times. We also examine how the approach back in time to the singularity is radically altered; happening much more abruptly, as a function of density, in an anisotropic universe. We also mention improved bounds that can arise from measurements of primordial gravitons, Weakly interacting massive particles, and neutrinos.
\end{abstract}

\maketitle


\section{Introduction}

In standard treatments of cosmology, one assumes statistical isotropy. Such an assumption seems well justified from various measurements in cosmology. In this work, our goal is to provide more precise observational bounds on global anisotropy. By global anisotropy, we mean a form in which the metric is anisotropic even after averaging. 

In particular, we shall consider the case in which the Friedmann Lemaitre Robertson Walker (FLRW) metric is altered so that the three axes of space can each expand at a priori different rates. A first basic question is then: if one allows for different expansion rates in different directions; how does this evolve into the late universe? Here we first show that in fact the difference in expansion rate asymptotically vanishes into the far future. Therefore any initial global anisotropies (of this form) are absent in the late universe.  We compute precisely the rate at which these vanish for both matter and radiation dominated eras. 

Conversely, this global anisotropy becomes larger back in time into the early universe. We compute this growth and find strikingly different behavior compared to standard FLRW cosmology as we approach the Big Bang singularity. In particular, while two of the 3 axes approach zero size, the third (typically) approaches infinite size. This all happens as the local energy density approaches infinity. So the very early stage of the universe could have been a very unusual shape -- dramatically elongated along just one axis. That one axis then contracts, then expands, and all axes eventually expand at similar rates. This all happens within known physics (for example, there is no violation of the null energy condition). Early interesting work on anisotropic universes, includes Refs.~\cite{Lemaitre:1933gd,Jacobs:1968zz}. Further interesting developments include Refs.~\cite{Caderni:1979ek,Wald:1983ky,Juszkiewicz,Kamionkowski:1990ni,Maleknejad:2012as,Schucker:2014wca,Dhawan:2022lze,Cowell:2022ehf,Aluri:2022hzs,Krishnan:2022uar,Ebrahimian:2023svi,Dehpour:2023wyy,Campanelli:2006vb,Tedesco:2018dbn,Akarsu:2019pwn,Amirhashchi:2018bic,Akarsu:2021max,Yadav:2023yyb}. Often (though not all) in this literature, the focus has been on late time constraints on anisotropy. However, as we will show, for the class of models of interest here (Bianchi-Type I) the most interesting constraints come  from the very early universe. 

We use several observations to constrain this possible early anisotropic era. The fact that current data is broadly compatible with an isotropic universe means that this era, if it occurred at all, must have only been in the very early universe. We use constraints from the cosmic microwave background (CMB) and Big Bang nucleosynthesis (BBN) to place bounds on the time of this era. Furthermore, we consider improving these bounds considerably in the future if there are observations of Weakly interacting massive particles (WIMPs), primordial neutrinos, or primordial gravitons. 

The outline of our paper is as follows: 
In Section \ref{Anisotropy} we present the equations of an anisotropic universe. 
In Section \ref{Matter} we study a matter dominated era.
In Section \ref{Radiation} we study a radiation dominated era.
In Section \ref{CMBBounds} we discuss bounds from CMB and BBN. 
In Section \ref{Singularity} we discuss the approach to the Big Bang singularity.
In Section \ref{Thermal} we discuss thermal relics, including gravitons and WIMPs.
In Section \ref{Decoupled} we discuss anisotropy in the distribution of relic neutrinos. 
Finally, in Section \ref{Discussion} we conclude.

\section{Anisotropic Universe}\label{Anisotropy}

Let us consider a universe that is homogenous, but anisotropic. We shall focus on a spatially flat universe. In co-moving coordinates, the metric is taken to be
\beq
ds^2=-dt^2+a(t)^2dx^2+b(t)^2dy^2+c(t)^2dz^2
\eeq
where $a,\,b,\,c$ are the scale factors associated with the $x,\,y,\,z$ axes, respectively. This is a Bianchi Type-I form of metric. 

We are assuming standard general relativity $G_{\mu\nu}=8\pi G\,T_{\mu\nu}$ with standard sources. 
The $\{tt\}$ component of the Einstein field equations gives a modified first Friedmann equation as
\beq
{1\over3}\left(H_aH_b+H_aH_c+H_bH_c\right)={8\pi G\over 3}\rho,
\label{First}\eeq
where $\rho$ is energy density, and
 the Hubble parameters associated with each axis are defined as
\beq
H_a={\dot{a}\over a},\quad
H_b={\dot{b}\over b},\quad
H_c={\dot{c}\over c}.
\eeq
Eq.~(\ref{First}) is reminiscent of the usual first Friedmann equation, with the difference that the left hand side is replaced with a kind of average of the Hubble parameters squared.

The $\{xx\},\,\{yy\}$, and $\{zz\}$ components of the Einstein field equations are
\bea
H_bH_c+{\ddot{b}\over b}+{\ddot{c}\over c}&=&-8\pi G\,\pr_x,\\
H_aH_c+{\ddot{a}\over a}+{\ddot{c}\over c}&=&-8\pi G\,\pr_y,\\
H_aH_b+{\ddot{a}\over a}+{\ddot{b}\over b}&=&-8\pi G\,\pr_z,
\eea
where $\pr_x,\,\pr_y,\,\pr_z$ are the pressures in the $x,\,y,\,z$ directions, respectively.

We can combine the above 4 Einstein equations to obtain a modified second Friedmann equation  as
\beq
{1\over3}\left({\ddot{a}\over a}+{\ddot{b}\over b}+{\ddot{c}\over c}\right) = -{4\pi G\over 3}(\rho+\pr_x+\pr_y+\pr_z)
\eeq
which is reminiscent of the usual second Friedmann equation, with the difference that the left hand side is replaced with a kind of average of the acceleration parameters.

Relatedly, we can form the continuity equation, as follows
\beq
\dot{\rho} = -\left(H_a(\rho+\pr_x)+H_b(\rho+\pr_y)+H_c(\rho+\pr_z)\right),
\eeq
which generalizes the usual continuity equation of $\dot{\rho}=-3H(\rho+\pr)$.

\section{Matter Domination}\label{Matter}

In a matter dominated era, we have $\pr_x=\pr_y=\pr_z=0$. The above continuity equation says that the energy density evolves as
\beq
\rho(t)=\rho(t_0){a(t_0)b(t_0)c(t_0)\over a(t) b(t) c(t)},
\eeq
where $t_0$ is some reference time. This makes good physical sense; the energy density of matter evolves as
$\rho(t)\propto 1/V_\text{phys}$, with $V_\text{phys}\propto a\,b\,c$, the physical volume.

In this case, the spatial parts of the Einstein equations are relatively simple:
\bea
H_bH_c+{\ddot{b}\over b}+{\ddot{c}\over c}=0,\label{GxxMatter}\\
H_aH_c+{\ddot{a}\over a}+{\ddot{c}\over c}=0,\label{GyyMatter}\\
H_aH_b+{\ddot{a}\over a}+{\ddot{b}\over b}=0.\label{GzzMatter}
\eea

\subsection{Axisymmetric}
We can find an exact solution of these equations if we consider the special case in which 2 of the 3 scale factors are equal. Let us set $a=b$, but allow $c$ to be different. Then the metric is axisymmetric around the $z$-axis. Eq.~(\ref{GzzMatter}) becomes very simple
\beq
H_a^2+2\,{\ddot{a}\over a}=0.
\eeq
In this limit $a$ is not mixed with the other scale factor $c$, and so it has a standard solution for a matter dominated era
\beq
a(t)=a_0\left(t-\tau\over t_0\right)^{2/3},
\eeq
(where $a_0,\,t_0,\,\tau$ are constants.)
Inserting this into Eq.~(\ref{GyyMatter}), we can then solve for $c(t)$, finding the general solution
\beq
c(t) = c_0\left(t-\tau\over t_0\right)^{2/3}\left(1+\tilde{c}\left(t_0\over t-\tau\right)\right).
\eeq
Here $\tilde{c}$ is the key new parameter of the solutions, which measures the amount of anisotropy. It we set $\tilde{c}=0$, then we return to an isotropic universe. For $\tilde{c}\neq 0$, the universe is anisotropic.

We note that even with $\tilde{c}\neq 0$, at late times $c(t)\to c_0(t/t_0)^{2/3}$, which is the standard relation in a matter dominated universe. So the solution asymptotes to an isotropic universe at late times. 
In fact we can set $c_0=a_0$ without loss of generality, by a rescaling of $x,\,y,\,z$, making it clear that the metric is FLRW at late times. 

\begin{figure}[t!]
\centering
\includegraphics[width=0.865\columnwidth]{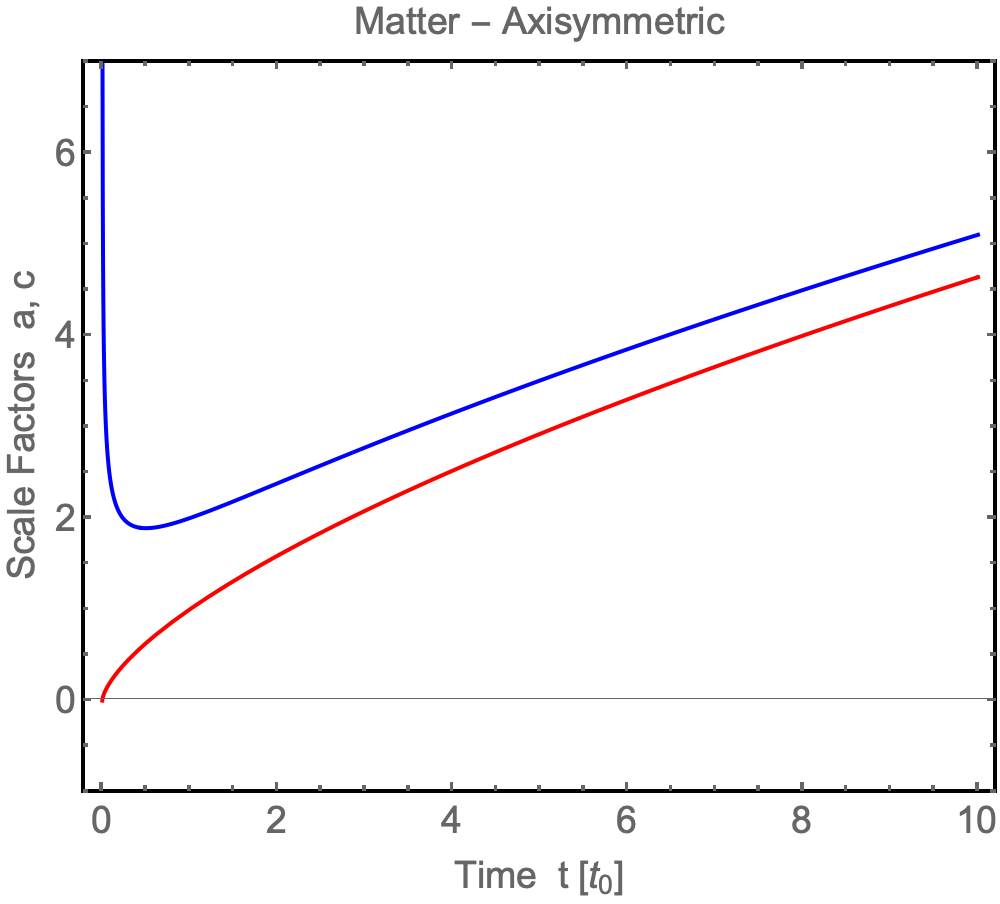}
\includegraphics[width=0.88\columnwidth]{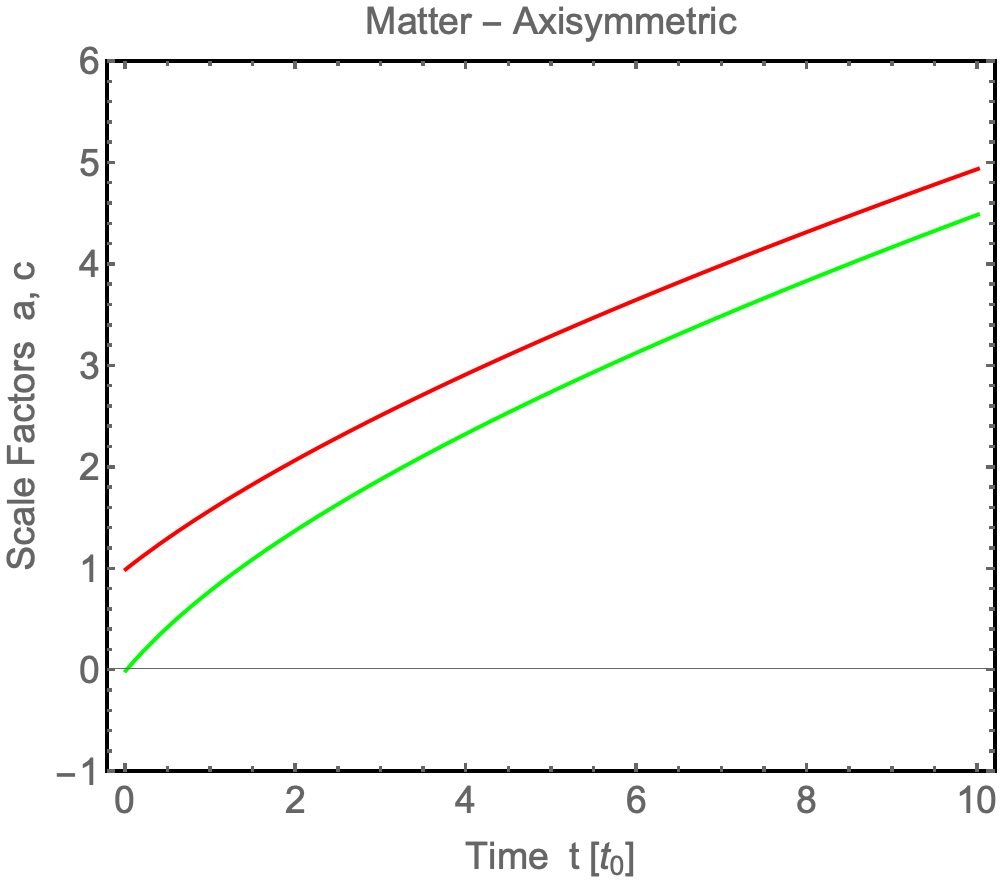}
\includegraphics[width=0.88\columnwidth]{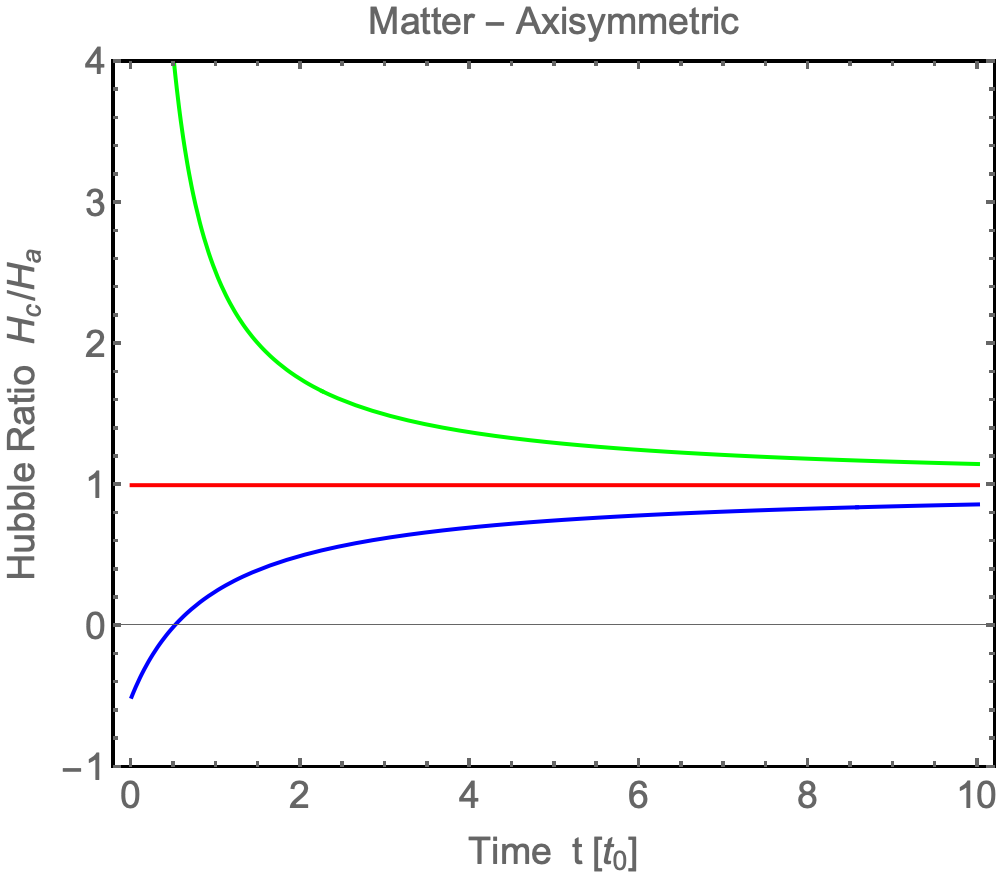}
\caption{Evolution of universe for a matter dominated, axisymmetric universe ($a=b$). 
Red curve is $a(t)$. Top panel is $\tilde{c}=+1$ with blue curve $c(t)$. Middle panel is $\tilde{c}=-1$ with green curve $c(t)$.
Bottom panel is Hubble ratio $H_c/H_a$ for each of the above cases ($\tilde{c}=+1$ in blue and $\tilde{c}=-1$ in green).} 
\label{FigMatterAxi} 
\end{figure}

Conversely, the universe could be very anisotropic at early times. A priori we cannot say what sign of $\tilde{c}$ should be taken. We could have $\tilde{c}>0$, in which case the scale factor along the $z$-axis goes to infinity as $t\to\tau$ (see top panel of Figure \ref{FigMatterAxi}, where we set $\tau=0$). Or we could have $\tilde{c}<0$, in which case the scale factor along the $z$-axis goes to zero at a time when the scale factor along the other axes remains finite  (see middle panel of Figure \ref{FigMatterAxi}, where we shift $\tau$ to bring the singularity to $t=0$). 

The associated Hubble parameters are
\bea
&&H_a=H_b={2\over 3(t-\tau)},\\
&&H_c={2\over 3(t-\tau)}\left(1-\tilde{c}(t_0/(2(t-\tau)))\over1+\tilde{c}(t_0/(t-\tau))\right).
\eea
The ratio of these Hubble parameters is given in bottom panel of Figure \ref{FigMatterAxi}.
At late times, these Hubble parameters approach each other as
\beq
H_c=H_a\left(1-{3\tilde{c}\over2}\left(t_0\over t\right)+\ldots\right).
\eeq
So the corrections decrease as $1/t$. 

\subsection{Fully Asymmetric}

\begin{figure}[t!]
\centering
\includegraphics[width=0.88\columnwidth]{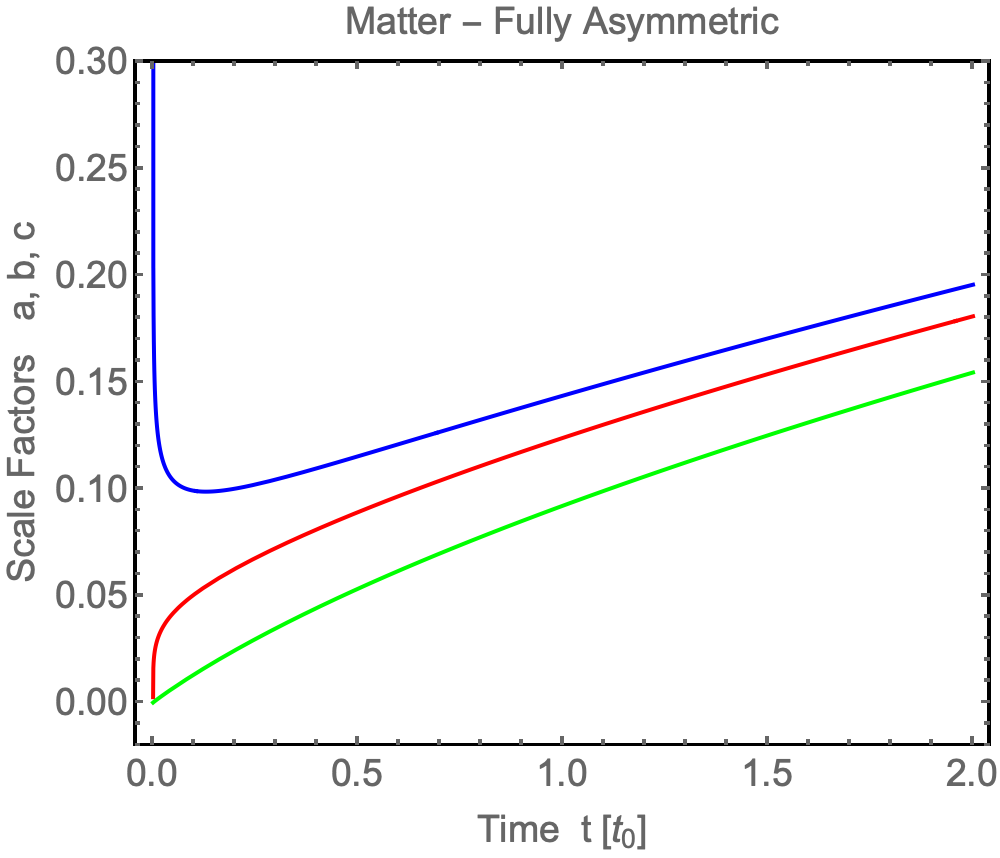}
\includegraphics[width=0.88\columnwidth]{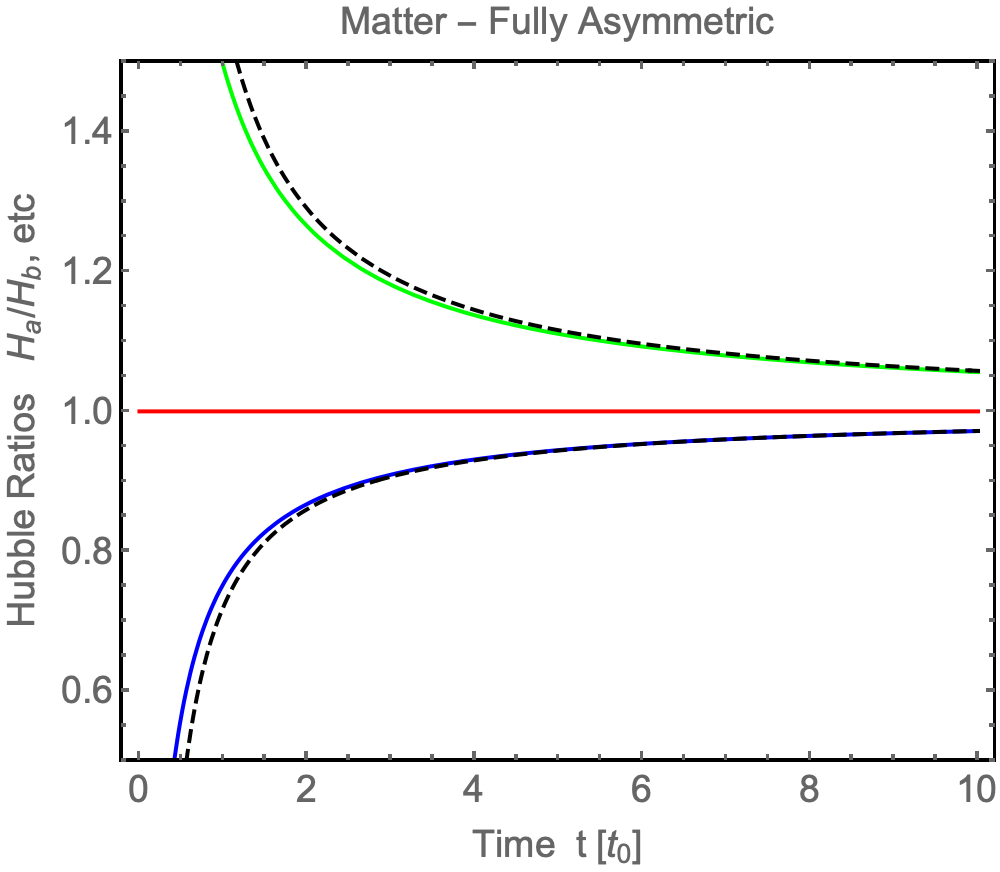}
\includegraphics[width=0.88\columnwidth]{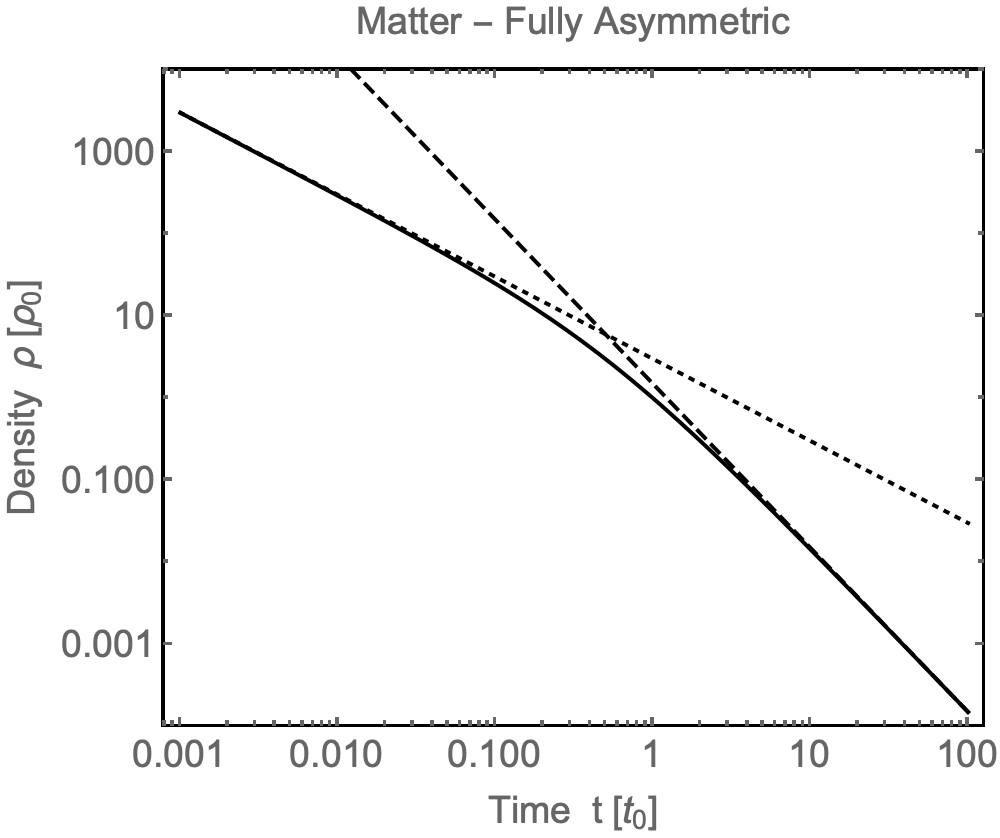}
\caption{Evolution of universe for a matter dominated, fully asymmetric universe ($a\neq b\neq c$). 
Blue curve is $a(t)$, red curve is $b(t)$, and green curve is $c(t)$. Middle panel are the Hubble ratios $H_a/H_b$ in blue and $H_c/H_b$ in green (with $H_b/H_b=1$ in red). We chose an initial condition with $H_a(t_0)<H_b(t_0)$ and $H_c(t_0)>H_b(t_0)$. The dashed curves are the late time asymptotes of Eqs.~(\ref{MatterAsymptote1}--\ref{MatterAsymptote3}). Bottom panel shows density $\rho$.}
\label{FigMatterDiff} 
\end{figure}

Now let us consider the more general case in which all the scale factors $a,\,b,\,c$ are different.  In this case, we do not have an analytical solution of the above equations, since the equations are all coupled. Nevertheless, we can solve the equations numerically. Our results are given in Figure \ref{FigMatterDiff}.

At late times, we can perform a series expansion to obtain the form of the solutions. We insert the following
\bea
a(t)=a_0\left(t\over t_0\right)^{2/3}\!\left(1+\tilde{a}_1\left(t_0\over t\right)+\tilde{a}_2\left(t_0\over t\right)^{\!2}+\ldots\right)&&\,\,\,\,\,\,\,\,\,\,\label{MatterAsymptote1}\\
b(t)=b_0\left(t\over t_0\right)^{2/3}\!\left(1+\tilde{b}_1\left(t_0\over t\right)+\tilde{b}_2\left(t_0\over t\right)^{\!2}+\ldots\right)&&\label{MatterAsymptote2}\\
c(t)=c_0\left(t\over t_0\right)^{2/3}\!\left(1+\tilde{c}_1\left(t_0\over t\right)+\tilde{c}_2\left(t_0\over t\right)^{\!2}+\ldots\right)&&\label{MatterAsymptote3}
\eea
We can always set $a_0=b_0=c_0$, without loss of generality, so that the scale factors match at late times. The coefficients, $\tilde{a}_1,\,\tilde{b}_1,\,\tilde{c}_1$ parameterize the deviation from isotropy. However, not all 3 of these parameters are meaningful, as we can always perform a shift on time $t$ to map one of them to zero. For example, we can transform $t\to t-3\,\tilde{b}_1\,t_0/2$ to eliminate $\tilde{b}_1$. So we in fact only have 2 residual anisotropy parameters ($\tilde{a}_1$ and $\tilde{c}_1$, say), which is obviously the correct amount to describe the {\em ratio} of Hubble parameters $H_a/H_b$, $H_c/H_b$. The higher order coefficients $\tilde{a}_2,\,\tilde{b}_2,\,\tilde{c}_2$, etc,  are all computable from insertion into the Einstein equations and matching order by order in an expansion; so they do not introduce further parameters.

By truncating the expansion at just the leading $\mathcal{O}(t^{2/3})$ and subleading $\mathcal{O}(t^{-1/3})$ terms for the Hubble parameters, we obtain the dashed curves of Figure \ref{FigMatterDiff}. We see that they match the full numerical result rather  well at late times.

\section{Radiation Domination}\label{Radiation}

In a radiation dominated era, we have to include pressure. Since we are allowing for an anisotropic universe, we could imagine that the pressures in the different directions $\pr_x,\,\pr_y,\,\pr_z$ are different.  However, so long as the material undergoes sufficiently rapid interactions, it is expected to have an isotropic pressure $\pr=\pr_x=\pr_y=\pr_z$. We shall focus on this situation in this section. For the case in which interactions are slow, especially when a species decouples, then there can be different free streaming in different directions, leading to different pressures. We will examine this in more detail in Section \ref{Decoupled}.

Assuming isotropy in pressure, then we have the standard relation for radiation 
\beq
\pr={\rho\over3}.
\eeq
The continuity equation then gives the following time dependence of energy density
\beq
\rho(t)=\rho(t_i)\left(a(t_i)b(t_i)c(t_i)\over a(t) b(t) c(t)\right)^{4/3}.
\eeq

In this radiation era, the equations are moderately more complicated than the matter era. Even when one considers the axisymmetric case with $a=b$, the Einstein equations now still have $a$ and $b$ coupled through the pressure. Hence we do not have exact solutions.

\begin{figure}[t!]
\centering
\includegraphics[width=0.87\columnwidth]{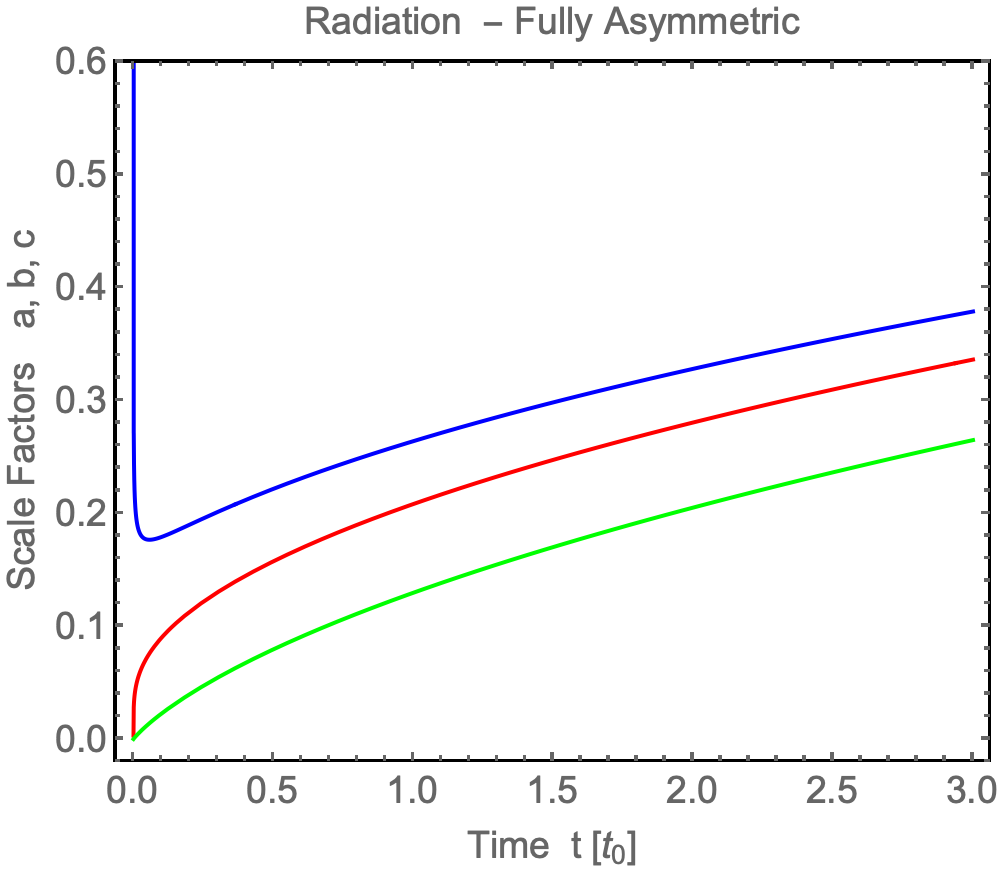}
\includegraphics[width=0.87\columnwidth]{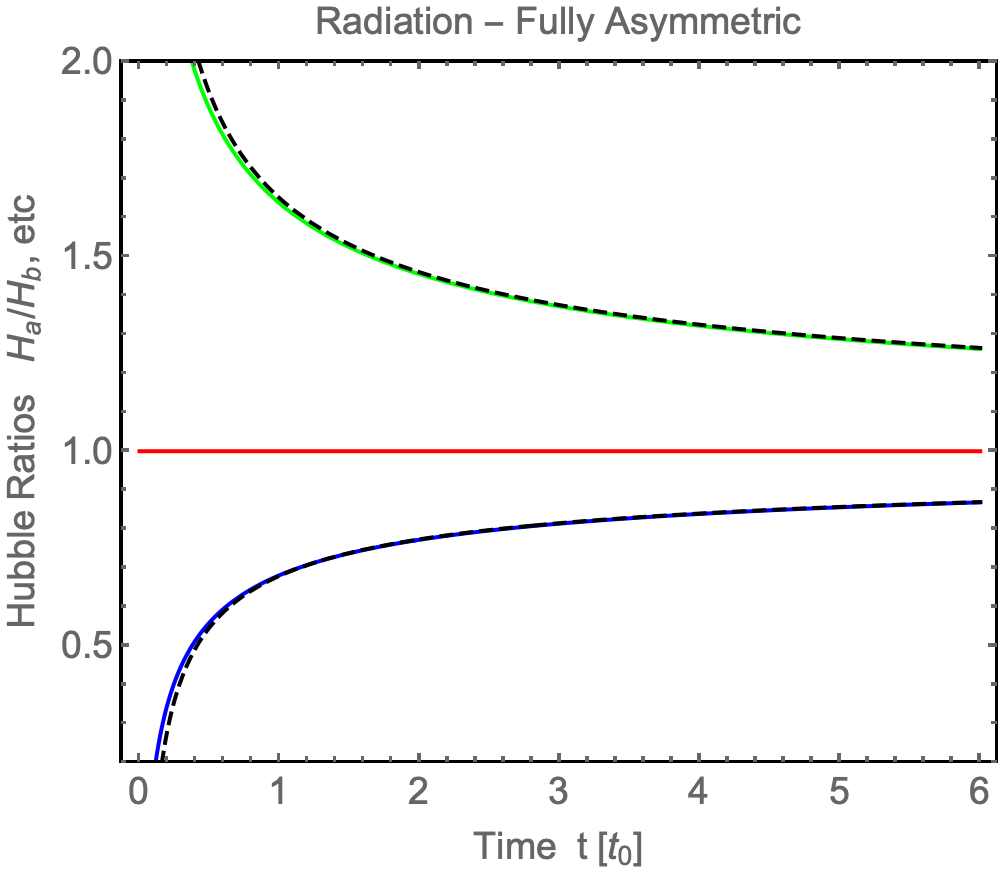}
\includegraphics[width=0.87\columnwidth]{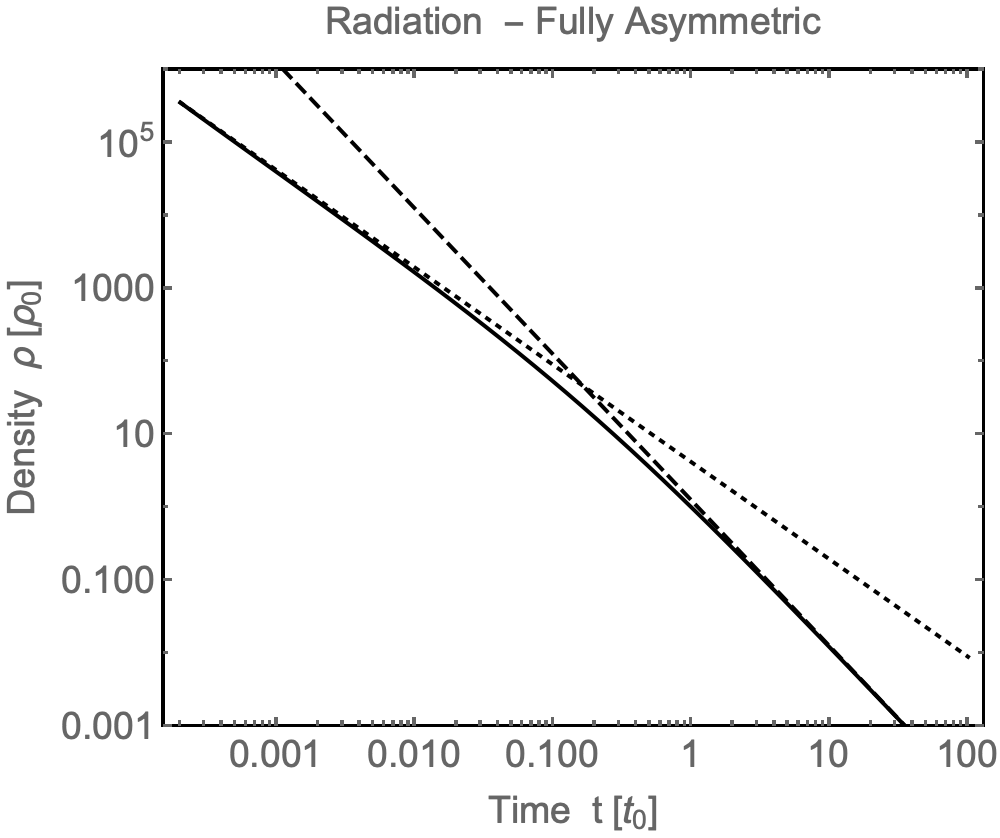}
\caption{Evolution of universe for a radiation dominated, fully asymmetric universe ($a\neq b\neq c$). 
Blue curve is $a(t)$, red curve is $b(t)$, and green curve is $c(t)$. Middle panel are the Hubble ratios $H_a/H_b$ in blue and $H_c/H_b$ in green (with $H_b/H_b=1$ in red). We chose an initial condition with $H_a(t_0)<H_b(t_0)$ and $H_c(t_0)>H_b(t_0)$. The dashed curves are the late time asymptotes of Eqs.~(\ref{RadiationAsymptote1}--\ref{RadiationAsymptote3}). Bottom panel shows density $\rho$.}
\label{FigRadiationDiff} 
\end{figure}

Nevertheless, we can still solve the Einstein equations numerically. Our results  are given in Figure \ref{FigRadiationDiff}. 

Once again, we can perform a late time series expansion. In this case, the series requires fractional powers as
\bea
a(t)=a_0\left(t\over t_0\right)^{1/2}\!\left(1+\tilde{a}_1\left(t_0\over t\right)^{\!1/2}+\tilde{a}_2\left(t_0\over t\right)+\ldots\right)&&\,\,\,\,\,\,\,\,\,\,\label{RadiationAsymptote1}\\
b(t)=b_0\left(t\over t_0\right)^{1/2}\!\left(1+\tilde{b}_1\left(t_0\over t\right)^{\!1/2}+\tilde{b}_2\left(t_0\over t\right)+\ldots\right)&&\label{RadiationAsymptote2}\\
c(t)=c_0\left(t\over t_0\right)^{1/2}\!\left(1+\tilde{c}_1\left(t_0\over t\right)^{\!1/2}+\tilde{c}_2\left(t_0\over t\right)+\ldots\right)&&\label{RadiationAsymptote3}
\eea
By insertion into the Einstein equations, we find that the leading coefficients, must obey the relation
\beq
\tilde{a}_1+\tilde{b}_1+\tilde{c}_1=0.
\eeq
This again makes good physical sense: it cuts us down to 2 residual parameters, say $\tilde{a}_1,\,\tilde{c}_1$ (with $\tilde{b}_1=-\tilde{a}_1-\tilde{c}_1$) describing the anisotropy. Then all other higher order parameters, $\tilde{a}_2,\,\tilde{b}_2,\,\tilde{c}_2$, etc, are computable from these parameters from solving the Einstein equations order by order (similar to the matter era case described earlier). 

\section{CMB and BBN Bounds}\label{CMBBounds}

We first consider the cosmic microwave background (CMB). This is well measured to be consistent with statistical isotropy; with variations in temperature at the $\delta T/T\sim 10^{-5}$ level. If at the time of CMB (redshift $z\sim 1100$) the universe carried significant  global anisotropy of the sort analyzed here, this would distort the CMB observations across the sky. By imposing that the anisotropy in the Hubble parameters was no more than $10^{-5}$ we can place a bound
\beq
\delta_{ac}\equiv{2|H_a-H_c| \over H_a+H_c}\lesssim 10^{-5}\quad\mbox{at}\,\,\,z\sim 1100
\eeq
and similarly for the pairs $H_a,\,H_b$, etc. This is a very tiny anisotropy at this redshift $z\sim 1100$. 

Let us suppose this anisotropy takes on some value at CMB $\delta(\CMB)$, such as $\delta(\CMB)=10^{-5}$ to saturate the above bound. We can ask: at what earlier time was the anisotropy $\mathcal{O}(1)$? To determine this, we should do this in 2 steps: (i) evolve back from the CMB at a time $t_{\tCMB}\approx 380,000$\,yrs and matter dominated to the time of matter-radiation equality $t_{\teq}\approx 70,000$\,yrs. (ii) evolve back from equality and radiation dominated to the time of the $\mathcal{O}(1)$ anisotropy. 
During matter domination, the anisotropy evolves as $\sim 1/t$ and during radiation the anisotropy evolves as $\sim 1/\sqrt{t}$. During the first step, we have
\beq
\delta(\eq)\approx\delta(\CMB)\,\left(t_{\tCMB}\over t_{\teq}\right)\approx 5.4\,\delta(\CMB)
\eeq
During the second step, we have
\beq
\delta(t)\approx\delta(\eq)\,\left(t_{\teq}\over t\right)^{1/2}
\eeq
The time $t_A$ at which the anisotropy is $\mathcal{O}(1)$, i.e., $\delta(t_A)= 1$ can be readily obtained as
\beq
t_A\approx 30\,t_{\teq}\,\delta(\CMB)^2
\eeq
So if we saturate the CMB level of anisotropy $\delta(\CMB)=10^{-5}$, this time is
\beq
\delta(\CMB)\approx 10^{-5}\quad \leftrightarrow\quad t_A\approx 100\,\mbox{minutes}
\eeq
This time is somewhat later than Big Bang nucleosynthesis (BBN) and so it would cause huge disruption from the standard cosmology. If we impose a more conservative value of $\delta(\CMB)=10^{-6}$, so to be 0.1 times the value of the known anisotropies, this makes $t_A\approx 1$\,minute, which is in fact around the start of BBN. 

So as to not significantly alter the predictions of BBN, including its successful predictions of the relic helium, deuterium, and (partially successful) lithium abundances, we should expect a relatively small amount of anisotropy at that time. This implies a slightly sharper bound. For the anisotropy to be  $\lesssim10\%$ at the time of 1 minute (the beginning of BBN), so as to be compatible with successes of BBN, we need
\beq
\delta(\CMB)\lesssim 10^{-7}\quad \leftrightarrow\quad  t_A\lesssim  0.6\,\mbox{seconds}
\eeq
Hence this pushes the anisotropy timescale to less than a second after the Big Bang, which is  a temperature of $T\sim$\,MeV, around the time of neutrino decoupling and the subsequent electron-position annihilation.
In this case the first second of the universe is still radically altered, a point we now turn to.

\section{Towards the Singularity}\label{Singularity}

Let us now examine how the anisotropy alters the very early universe as we head back towards the singularity. 

In both a matter dominated era and a radiation dominated era (the latter being more relevant for the very early universe, while the former could be relevant in some models that prefer an even earlier matter era), we derive a different expansion compared to that given above. As $t\to0$ one can show that power law solutions apply, which we write as
\beq
a(t)\propto t^\alpha,\quad b(t)\propto t^\beta,\quad c(t)\propto t^\gamma,
\label{exponents}\eeq
for some exponents $\alpha,\,\beta,\,\gamma$.  By inserting this into the Einstein equations and then taking the small $t$ limit, we can establish the allowed values these exponents. 

In any era with anisotropy, we find that these exponents are related by the following pair of conditions:
\bea
\alpha\beta+\alpha\gamma+\beta\gamma=0,\\
\alpha+\beta+\gamma=1.
\eea
We can use this pair of conditions to solve for two of the exponents in favor of the other one. So for example, we can express $\alpha$ and $\gamma$ in terms of $\beta$ as
\bea
\alpha&=&{1\over2}(1-\beta\mp\sqrt{1+2\beta-3\beta^2}),\\
\gamma&=&{1\over2}(1-\beta\pm\sqrt{1+2\beta-3\beta^2}),
\eea
where the $\mp$ choice for $\alpha$ requires the corresponding $\pm$ choice for $\gamma$. 
Special cases of this are (i) $\alpha=\beta=2/3$, and $\gamma=-1/3$, or (ii) $\alpha=\beta=0$, and $\gamma=1$. These correspond to the early time limit of the axisymmetric cases studied earlier. While more general values of $\beta$ lead to fully anisotropic behavior in all 3 axes. In particular, the generic case is that two of the 3 exponents are positive and one is negative. This corresponds to two axes approaching zero size and the other approaching infinite size, as seen in the top panel of Figure \ref{FigMatterDiff} and \ref{FigRadiationDiff}. We note that this can have important consequences for theories of the very early universe, such as inflation. However, the full ramifications for theories of the beginning of the universe is beyond the scope of the present work. We have checked that vacuum domination tends to erase anisotropies into the future, but we shall not go into the details here. The time at which significant anisotropy can occur shall be addressed in the next subsection.

Note that since the sum of the exponents is always
$\alpha+\beta+\gamma=1$, we always have number density scaling as $n\propto1/(abc)\propto 1/t$. So we get a different scaling of densities compared to the usual FLRW case.
In a matter era, this means the energy density scales as
\beq
\rho(t)\propto {1\over a(t)b(t)c(t)}\propto {1\over t}\quad\mbox{(matter)},
\eeq
as $t\to0$. While in a radiation era, this means the energy density scales as
\beq
\rho(t)\propto {1\over (a(t)b(t)c(t))^{4/3}}\propto {1\over t^{4/3}}\quad\mbox{(radiation)},
\eeq
as $t\to 0$. 
 Let us contrast this with the standard result in an isotropic FLRW universe with either matter or radiation domination
\beq
\rho_\text{FLRW}(t)\propto{1\over t^2}.
\eeq
Hence the approach of the density to the singularity is less steep than in the isotropic case. 
The density $\rho$  versus time is given in the bottom panel of Figures \ref{FigMatterDiff} (matter) and \ref{FigRadiationDiff} (radiation). 
For the matter case, the early time asymptote of $\rho\propto 1/t$ is the dotted line of Figure \ref{FigMatterDiff}. 
For the radiation case, the early time asymptote of $\rho\propto 1/t^{4/3}$ is the dotted line of Figure \ref{FigRadiationDiff}. 

At late times, the system isotropizes and we see in the figures how the density then scales in the usual way, i.e., $\rho\to\rho_\text{FLRW}\propto 1/t^2$; as given by the dashed lines.

\subsection{Timescale}

As above, let us denote $t_A$ the characteristic timescale at which the anisotropy is $\mathcal{O}(1)$. Then we can summarize the above by writing
\beq
\rho(t)=\rho_\text{FLRW}(t)\times\Bigg{\{}\begin{array}{c}(t/t_A)^p,\quad t\lesssim t_A\\1,\quad \quad\quad\,\, t\gtrsim t_A\end{array}
\eeq
where $p=1$ for matter and $p=2/3$ for radiation. 

Now recall that the usual FLRW equation is
\beq
\rho_\text{FLRW}=3\,\mpl^2\,H^2,
\eeq
where $\mpl=1/\sqrt{8\pi G}$ is the (reduced) Planck mass. Recall that the Hubble parameter is $H=2/(3t)$ for matter and $H=1/(2t)$ for radiation. Suppose there is some interesting phenomenon at some density $\rho_*$, such as a phase transition (QCD or electroweak) or BBN. In an isotropic FLRW universe, the corresponding time $t_*$ at which this occurs is 
\beq
t_{*\text{FLRW}}\sim {\mpl\over\sqrt{\rho_*}}.
\eeq

Now let us consider an anisotropic universe, and suppose the interesting phenomenon occurs at a time $t_*\lesssim t_A$, i.e., during the anisotropic phase. Using the above results, this is related to the density by
\bea
&& t_*\sim{\mpl^2\over\rho_*\,t_A}\sim{\mpl\sqrt{\rho_A}\over\rho_*}\quad\mbox{(matter)},\label{timematter}\\
&& t_*\sim{\mpl^{3/2}\over\rho_*^{3/4}\sqrt{t_A}}\sim{\mpl\,\rho_A^{1/4}\over\rho_*^{3/4}}\quad\mbox{(radiation)},\label{timerad}
\eea
where in the second step of the estimate, we used the fact that at $t_A$ we can roughly use the FLRW equation $t_A\sim\mpl/\sqrt{\rho_A}$. 
For an event in the very early universe of high energy density $\rho_*\gg\rho_A$, we have $t_*$ much less than the FLRW prediction. We return to this in Section \ref{Gravitons}.

\section{Thermal Relics}\label{Thermal}

For a species $X$ with number density $n_X$, the evolution of its number density is governed by the Boltzmann equation. For an anisotropic universe, the usual form is modified to
\beq
\dot{n}_X+(H_a+H_b+H_c)n_X=-\langle\sigma|v|\rangle(n_X^2-n_{X,eq}^2),
\eeq
where $n_{X,eq}$ is the equilibrium distribution.
The condition to maintain equilibrium is that the annihilation rate $\Gamma=\langle\sigma|v|\rangle n_X$ needs to be larger than the (average) Hubble rate 
\beq
H_\text{ave}={H_a+H_b+H_c\over3}.
\eeq
In the anisotropic era in the very early universe, with $t\ll t_A$, we have 
\beq
H_\text{ave}={\alpha+\beta+\gamma\over 3\,t}={1\over 3\,t}.\label{HaveV}
\eeq
(Compare this to the usual FLRW case during a radiation era of $H=1/(2t)$.)
Thus we recover a familiar rule to maintain equilibrium at time $t$ is $\Gamma\gtrsim 1/t$ (the precise boundary is only an $\mathcal{O}(1)$ number different than the usual FLRW case.)


\subsection{Primordial Gravitons}\label{Gravitons}

Suppose we have $\mathcal{O}(1)$ anisotropy at some temperature $T_A$, with $\rho_A\sim T_A^4$, where we are assuming an early radiation era. We can wonder: at what time is the Planck density era where $\rho_*\sim\mpl^4$? The usual answer is of course $t_{*\text{FLRW}}\sim 1/\mpl=\tpl$. But for an anisotropic universe, Eq.~(\ref{timerad}) gives
\bea
&& t_*\sim{T_A\over\mpl^2}\quad\mbox{(at Planck density)}.
\eea
For any temperature of anisotropy $T_A$ less than the Planck temperature, these timescales are shorter than the Planck time, and therefore we anticipate they have no physical meaning. This means that the Planck density era was so short lived that gravitons likely did not thermalize. 

We can be more precise about this as follows: At temperature $T$, the graviton-graviton annhilation rate into the Standard Model plasma is
\beq
\Gamma\sim{T^5\over\mpl^4}.
\eeq
From Eqs.~(\ref{timerad},\,\ref{HaveV}) the corresponding (average) Hubble rate at temperature $T$ was
\beq
H_\text{ave}\sim{T^3\over\mpl T_A}.
\label{HT}\eeq
Equating the above pair of results, $\Gamma\sim H_\text{ave}$, gives the temperature of graviton freeze-out $T_F$ of
\beq
T_F\sim{\mpl^{3/2}\over T_A^{1/2}}.
\eeq
For any $T_A$ less than the Planck temperature, $T_F$ is larger than the Planck temperature, which is likely unphysical.

This suggests there would be no relic graviton bath. This differs from the FLRW prediction of a relic species of gravitons from the Planck era (unless inflation intervened) \cite{Vagnozzi:2022qmc}. 
Therefore any future detection of primordial gravitons would rule out an anisotropic universe all the way back to the Planck era.

Alternatively, we can ask: Suppose physics breaks down for sub-Planckian time scales $t\lesssim t_*=\tpl$, then what is the corresponding density $\rho_*$? From Eq.~(\ref{timerad}) we obtain
\beq
\rho_*\sim T_A^{4/3}\mpl^{8/3} \quad\mbox{(at Planck time)}.
\eeq
So, for example, if there was an $\mathcal{O}(1)$ amount of anisotropy at $T_A\sim$ MeV, the era of electron-position annihilation, then this Planck time era has a density of
$\rho_*\sim(10^{11}\,\mbox{GeV})^4$. This density is lower than that required in typical theories of unification and inflation models. Hence this could have profound consequences.

\subsection{Thermal WIMPs}

Consider a massive particle in standard FLRW cosmology that was initially in thermal equilibrium then its annihilations into the Standard Model particles froze out. We shall generically refer to this as a ``WIMP" (irrespective of whether its interactions involve the weak force or not). The relic abundance is computed from the Boltzmann equation to be
\beq
\xi_\text{FLRW}\equiv{\rho_\text{WIMP}\over n_\gamma} \sim {1\over\langle\sigma|v|\rangle\mpl},
\label{xiNormal}\eeq
where we are using a conveninent (late) time indepdent variable $\xi$ which is the ratio of energy density of the WIMP to the photon number density.
The observed relic dark matter abundance is $\Omega_{obs}\approx0.26$. This translates into the time independent parameter of $\xi_{obs}\approx 3$\,eV. 
So this needs $\langle\sigma|v|\rangle\sim 1/(20\,\mbox{TeV})^2$ to give the observed relic abundance.

 However, this analysis needs to be revisited if this freeze-out occurred during an initial anisotropic era. Repeating the analysis with the (average) Hubble estimate during an anisotropic radiation era of Eq.~(\ref{HT}) gives roughly (up to some mild log dependence)
\beq
\xi\equiv{\rho_\text{WIMP}\over n_\gamma} \sim {T_F\over\langle\sigma|v|\rangle\mpl\,T_A},
\label{xiNew}\eeq
where the freeze-out temperature is typically $T_F\sim m_X/20$,
where $m_X$ is the mass of the WIMP. Compared to the standard FLRW case, this is an {\em increased} relic abundance by a factor of $\sim T_F/T_A$. 

In order to not over-close the universe, one needs to have a correspondingly {\em higher} annihilation cross section. Such WIMPs would be even easier to detect. Given current constraints, which require the annihilation cross section to be low, we cannot have a huge ratio $T_F/T_A\gg 1$.  This means a future detection of WIMPs would constrain $T_F/T_A\lesssim 1$ and so the anisotropy would be pushed back to at or before the WIMP freeze-out era.

The allowed range of WIMP masses and couplings has been constrained by both direct detection, indirect detection, and collider experiments. A simple scaling for the cross section is $\langle\sigma v\rangle \sim g^4/(16\pi m_X^2)$ where $g$ is a coupling \cite{Bauer:2017qwy}. If we impose the unitarity bound $g^2\lesssim 4\pi$ then the mass $m_X$ is bounded to avoid over closure. The usual bound from Eq.~(\ref{xiNormal}) is $m_X\lesssim 50$\,TeV \cite{Bauer:2017qwy}. But in the presence of an early anisotropic era with Eq.~(\ref{xiNew}), this bound is altered to $m_X\lesssim 50\sqrt{T_A/T_F}$\,TeV, which is a tighter bound. Therefore if the WIMP is discovered with a mass close to 50\,TeV, then the era of anisotropy would be pushed back to $T_A\gtrsim T_F\approx m_X/20\approx2.5$\,TeV. Moreover, if a lighter WIMP is discovered with a corresponding coupling $g$ that matches the standard calculation, the era of anisotropy would be pushed back to the corresponding $T_A\gtrsim m_X/20$. For any reasonable WIMP, this would still be orders of magnitude higher in temperature than the previous bound from CMB/BBN of Section \ref{CMBBounds}. 

Alternatively, if there is a mismatch between the measured WIMP mass, coupling, and predicted abundance, then it may indicate that an early anisotropic era took place.

\section{Decoupled Species}\label{Decoupled}

When species decouple and free stream, the average velocity squared in the $j$th direction can be written as
\beq
\langle v_j^2\rangle={g\over n}\int \! {d^3p\over(2\pi)^3}\,n_\text{occ}(p)\,{p_{j}^2\over E_p^2},
\eeq
where $n_\text{occ}$ is occupancy number, $E_p=\sqrt{p^2+m^2}$, and the normalization is provided by the number density 
\beq
n=g\int \! {d^3p\over(2\pi)^3}\,n_\text{occ}(p),
\eeq
The momentum redshifts as
\beq
p_x={a_{d}\over a}\,p_{dx},\quad
p_y={b_{d}\over b}\,p_{dy},\quad
p_z={c_{d}\over c}\,p_{dz},\quad
\eeq
where the d subscripts indicate some reference value, which may be taken to be the value at the time of decoupling. 

\begin{figure}[t]
\centering
\includegraphics[width=0.9\columnwidth]{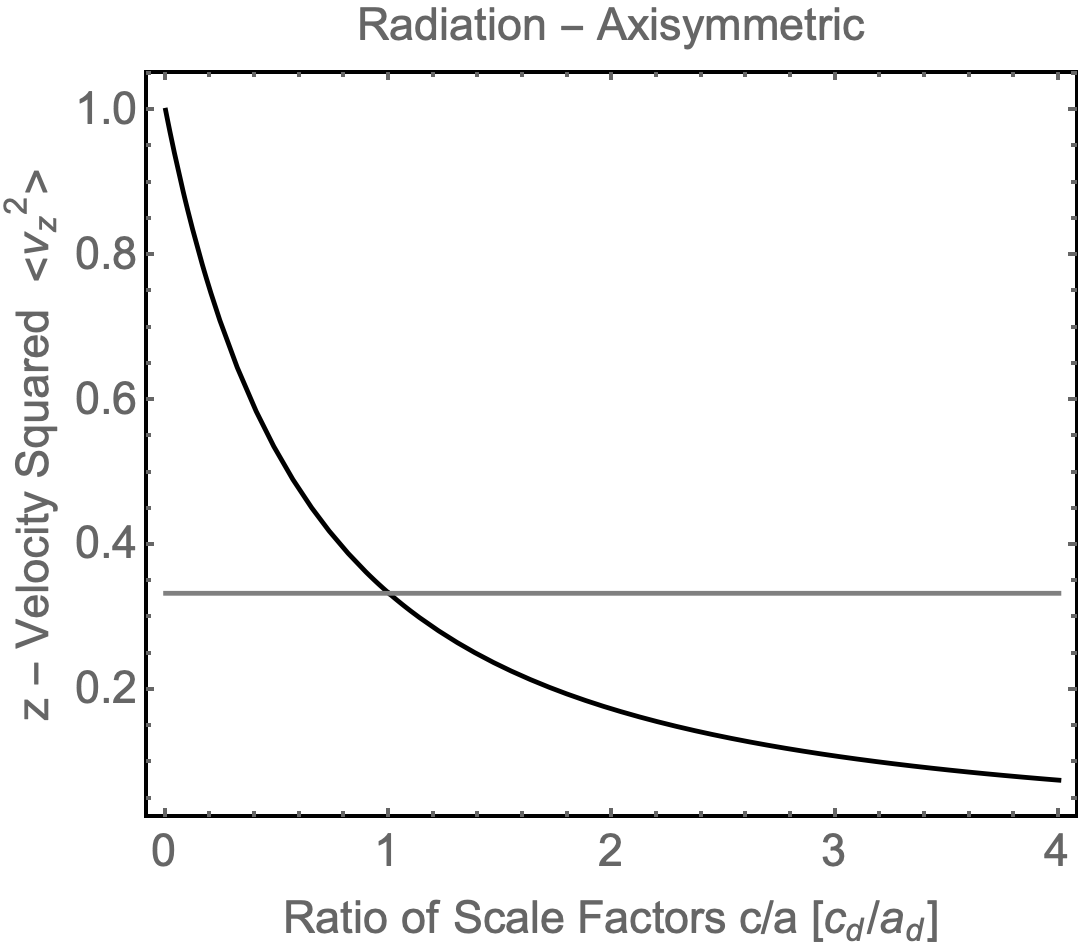}\\
\includegraphics[width=0.94\columnwidth]{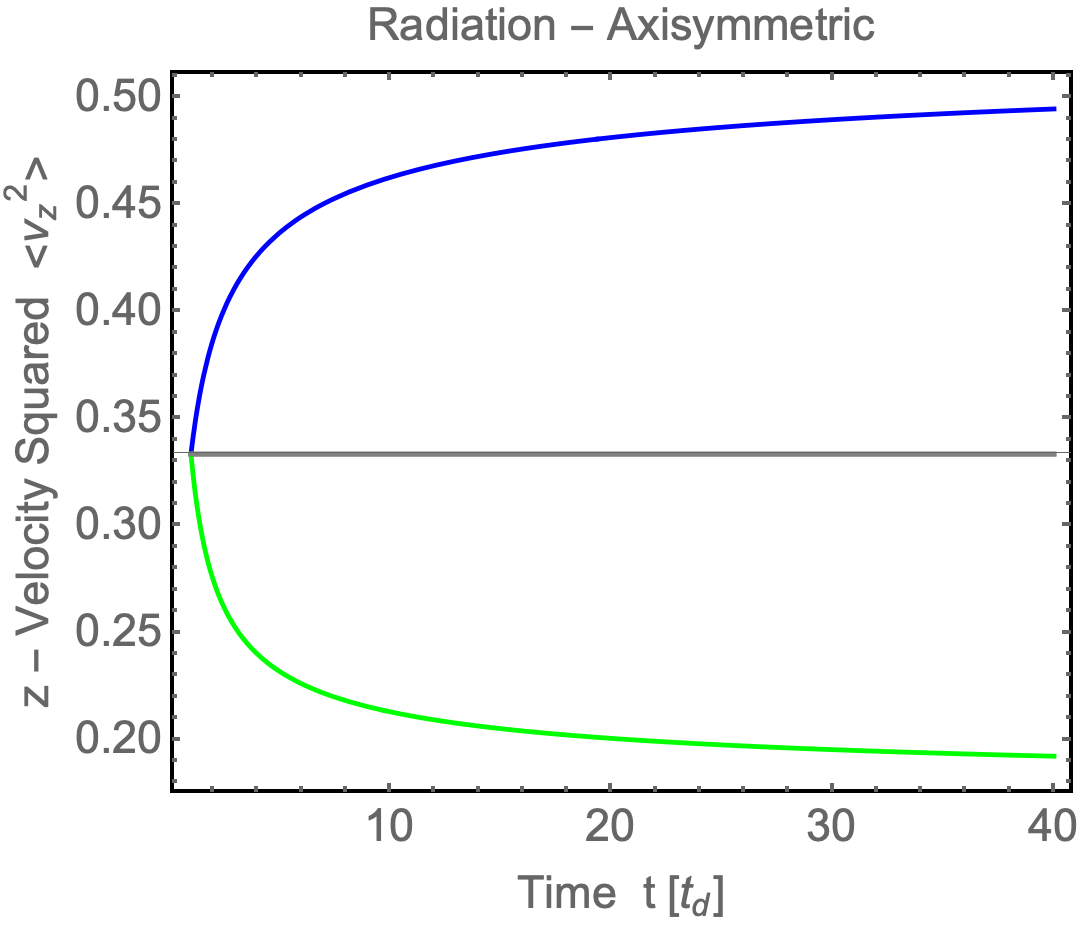}
\caption{Dependence of z-velocity squared $\langle v_z^2\rangle$ for a radiation dominated, axisymmetric universe  ($a= b$) after decoupling. Top panel gives this as a function of ratio of scale factors $c/a$. The grey horizontal line is the standard reference value of $1/3$. Bottom panel gives this as a function of time $t$; blue curve is for $c_d/a_d=2$ and green curve is for $c_d/a_d=1/2$.}
\label{FigVelSquared} 
\end{figure}

For the axisymmetric case in which $a=b$ (but $c$ is different) we have solved for this evolution. We assume that until decoupling we have an isotropic distribution $\langle v_x^2\rangle=\langle v_y^2\rangle=\langle v_z^2\rangle=1/3$ due to rapid interactions. However, it departs from this isotropy of velocities after decoupling. We then plot the average z-velocity squared versus scale factor and versus time in Figure \ref{FigVelSquared}. The blue curve is the case in which at the time of decoupling $c_d/a_d=2$ and the green curve is is the case in which at the time of decoupling $c_d/a_d=1/2$. We have used the leading asymptotic expansion described above for this plot. And we have ignored the particle mass here. In the first (second) case, we end up with faster (slower) velocities in the $z$-direction at late times; this follows because in this case the $z$ direction expands slower (faster), until the universe isotropizes at late times.

Applying the above to primordial neutrinos makes an interesting prediction: the speed of neutrinos in different directions will be different. This requires the detection of primordial neutrinos, which is an ongoing challenge.

\section{Conclusions}\label{Discussion}

In this work, we have considered a globally anisotropic universe.  In particular, we considered a Bianchi Type I model, in which the 3 axes of space can have different scale factors. We assessed this against a suite of different kinds of observables. Firstly, we showed that any initial anisotropy decreases forwards in time in an expanding universe; the relative difference between the axes Hubble parameters decreases as $1/\sqrt{t}$ in a radiation era and as $1/t$ in a matter era. This means that if there was such an anisotropic era, it would have occurred in the early universe. 

We used bounds from CMB to push back the anisotropic era to just minutes after the Big Bang. Moreover we used consistency of BBN to push the anisotropic era to just seconds after the Big Bang.

Then we computed the change in the behavior to the initial Big Bang singularity. It occurs much more abruptly than in standard FLRW. In particular, instead of Hubble depending on temperature as $H\sim T^2/\mpl$, we have that (average) Hubble depending on temperature as $H\sim T^3/(\mpl\,T_A)$, where $T_A$ is the temperature when the universe was $\mathcal{O}(1)$ anisotropic. Thus in the very early universe, for $T\gg T_A$, this is an especially large expansion rate. 

We showed that this means no primordial gravitons would have even been in thermal equilibrium, even at Planck densities. So a future detection of a relic thermal bath of gravitons would push the era of anisotropy all the way back to the Planck era. 

We then computed the change in the freeze out of a WIMP, finding that the relic abundance is enhanced relative to the usual formula by a factor of $\sim T/T_A$ as well. Therefore a future detection of WIMPs can probe this era and allow us to constrain, or infer the existence of, such an anisotropic phase. 

Finally, we mentioned a novel effect of relic decoupled species, such as neutrinos, acquiring different momenta along the different axes. Since this difference in momenta is injected at early times, it is not erased as the universe at large scales becomes isotropic at late times. This means that relic neutrinos today would have a residual anisotropic spectrum. 




\section*{Acknowledgments}
M.~P.~H.~ is supported in part by National Science Foundation grant PHY-2310572. 
A.~L.~ was supported in part by the Black Hole Initiative at Harvard University which is funded by grants from the John Templeton Foundation and the Gordon and Betty Moore Foundation.


\end{document}